\documentclass[10pt,letterpaper]{article}
\usepackage{amsmath}
\usepackage{opex3}
\usepackage{cite}

\usepackage{mathrsfs}
\usepackage{amsfonts}
\usepackage{amssymb}
\usepackage{eufrak}
\usepackage{multirow}

\begin{document}

\title{Hyper-parallel Toffoli gate on three-photon system with two degrees of freedom assisted by single-sided optical microcavities}

\author{Hai-Rui Wei$^{1,2}$* ,  Fu-Guo Deng$^{3}$  and  Gui Lu Long$^{1,2}$}

\address{$^{1}$ School of Mathematics and Physics, University of Science and Technology Beijing, Beijing 100083, China\\

$^{2}$ State Key Laboratory of Low-Dimensional Quantum Physics and Department of Physics, Tsinghua University, Beijing 100084, China\\

$^{3}$ Department of Physics, Applied Optics Beijing Area Major Laboratory, Beijing Normal University, Beijing 100875, China}

\email{*hrwei@ustb.edu.cn}


\begin{abstract}
Encoding qubits in multiple  degrees of freedom (DOFs) of a quantum system allows less-decoherence quantum information processing with much less quantum resources. We present a compact and scalable quantum circuit to determinately implement a hyper-parallel controlled-controlled-phase-flip (hyper-$\rm{C^2PF}$) gate on a three-photon system in both the polarization and spatial DOFs. In contrast with the one with many qubits encoding in one DOF only, our hyper-$\rm{C^2PF}$ gate operating two independent $\rm{C^2PF}$ gates on a three-photon system  with less decoherence, and reduces the quantum resources required in quantum information processing by a half. Additional photons, necessary for many approaches, are not required in the present scheme. Our calculation shows that this hyper-$\rm{C^2PF}$ gate is feasible in experiment.
\end{abstract}




\section{Introduction}\label{sec1}

Quantum computers promise to solve certain computational tasks which
are intractable for a classical one \cite{book}. Implementations
of universal quantum logic gates, the central requirement for a
quantum computer, have attracted widespread
attention \cite{synthesis-3CNOT,long1,long2,Controlled-Evolutions,Hybrid-architecture}.
Much was focused  on the two-qubit controlled-NOT (CNOT) gate or the
identical controlled-phase-flip (CFP) gate \cite{Atom-CPF1,Atom-CPF2}. Multiqubit gates play a
central role in quantum networks, quantum corrections, and quantum
algorithms, and they serve as a stepping stone for implementing a
scalable quantum computer. The simplest universal gate library in
multiqubit systems  is \{Toffoli  or Fredkin gate, Hadamard
gate\} \cite{Toffoli,Fredkin}. However, the inefficient
synthesis \cite{Tsing,Fredkincost} of Toffoli or Fredkin gate
increases the length and time of such two gates, and makes the gates
further susceptible to their environments. The minimized cost of a three-qubit Toffoli gate is five two-qubit gates \cite{Tsing}, and the decomposition of a generalized $n$-qubit Toffoli gate requires $O(n^2)$ two-qubit gates \cite{Universal}.  Some physical
architectures, including linear optics, superconducting circuits,
quantum dots (QDs), trapped ions, and diamond nitrogen-vacancy (NV) defect
centres, have been proposed to implement Toffoli (or the identical
$\rm{C^2PF}$) gate with one degree of freedom
(DOF) \cite{Toffoli-linear2,superconductgate1,Toffoli-superconduct1,superconductgate4,superconductgate5,weiSR,weioeQD,Toffoli-ion2,ourgatesNV}.

Quantum gates designed with minimized resources is crucial for
quantum computation. Two strategies are generally adopted to remedy
this problem: the one is to exploit a system with additional readily
accessible states (qudit) during the
computation \cite{additional2,additional4}; the second strategy is
to encode the information in multiple DOFs of a quantum system. That is
referred to as hyper-parallel quantum
gates \cite{renLPL,renSR,renhypercnotpra}, i.e., the quantum gates
operating  more than one independent operations simultaneously. By
the first approach, the quantum circuits for solid-state electronic CNOT
gate \cite{additional2} and flying photonic Toffoli
gate \cite{additional4} have been designed. By the second approach,
schemes for implementing hyper-parallel photon-based CNOT
gate \cite{renLPL,renSR,renhypercnotpra} and hyper-parallel
photon-matter-based universal gates \cite{wangtiejun1,wangtiejun2}
have been proposed. The hyperentangled-cluster-state-based quantum
computing have been demonstrated in recent
years \cite{hypercluster2}.

Single photon is one of the most popular candidates for quantum
information processing with  multiple DOFs because of  its
robustness against decoherence, exact  and flexible manipulation
with linear optics, and  many accessible qubitlike
DOFs \cite{photon-DOF2}, such as polarization, spatial, orbital
angular momentum, transverse, energy-time, time bin, and so on.
However, the photon-based  quantum computing in a small-scalable
fashion is a challenge in experiment because of the weak nonlinear
interactions at the single-photon level. Fortunately, this intrinsic
limitation can be partially solved by employing
linear-optics \cite{KLM} and completely solved by employing
cross-Kerr nonlinearity \cite{kerr1,sheng} or  photon-matter entangling
platforms \cite{renhypercnotpra,atom,platform1}. Up to now,
the giant Kerr nonlinearity is still a challenge in experiment.  The
mechanism of an emission-based photon-matter system can mediate a
required photon-photon or matter-matter interaction, and opens up
perspectives for directly  photon-based \cite{renLPL,renSR,weioe}, or
matter-based \cite{weiSR,weioeQD}  scalable quantum computing in a determinate way. In addition, it allows the non-destructive
photon-polarization (matter-spin) state measurement heralded by the
states of the matter (photonic) qubit.

The electronic spins associated with the NV centre stand out as an
attractive matter qubit because of their milliseconds coherence
time \cite{coherence1,coherence2} and stable single-photon emission
at room temperature, while can be manipulated on a subnanosecond time
scale \cite{manipulate4}. The exactly
initialization \cite{population2}, manipulation \cite{manipulate4},
and high-fidelity readout \cite{readout3} of an  NV centre electron
spin, which is crucial for quantum information processing, have been
reported in recent years.  Nowadays, the NV centres have been subject to
numerous applications: in 2007, Gurudev Dutt \emph{et
al}. \cite{distributed} demonstrated distributed quantum computing in
an NV centre. The Deutsch-Jozsa algorithm in a single NV centre was
demonstrated in 2010 \cite{algorithm}. In 2012, van der Sar \emph{et
al.} \cite{decoherence-protected} demonstrated the hybrid quantum
gates acting on the electron spin and the nearby nuclear spin. In 2014, Arroyo-Camejo \emph{et
al.} \cite{geonetric-single} and Zu \emph{et
al.} \cite{geonetric-CNOT}  demonstrated geometric single-qubit gates
and  CNOT gate within an NV centre, respectively. In recent years,
there have been a number of hallmark demonstrations of quantum
entanglement between an emitted single photon and a stored NV centre
electron spin (or between two separated NV centre spins) \cite{Togan,Kosaka,three-meter}.  Based on NV-centre-emission-based
entanglement, some interesting schemes for implementing universal
gates on NV centre electronic \cite{ourgatesNV} or photonic
qubits \cite{Wangchuan,ourphotongate}, and hyperentanglement
purification and concentration  were also proposed
recently \cite{ren}.

In this paper, we theoretically present an alternative scheme for
compactly implementing an optical hyper-$\rm{C^2PF}$ gate assisted
by  NV-centre-cavity systems. Our proposal has the following merits. First,
single photons encoded in both the polarization and  spatial DOFs, and
this strategy reduces the quantum resource by a half than that with many
qubits of one DOFs. Second, our universal gate works for a
three-photon system in a hyper-parallel way.  Third, additional
photons, necessary for cross-Kerr- or parity-check-based photonic
quantum computing, are not required. Fourth, our scheme is much simpler
than that one cascaded with single- and two-qubit (CNOT or CPF)
gates.

\section{Compact quantum circuit for implementing hyper-parallel $\rm{C^2PF}$ gate.} \label{sec2}

\begin{figure}[tpb]
\begin{center} \includegraphics[width=8.5 cm,angle=0]{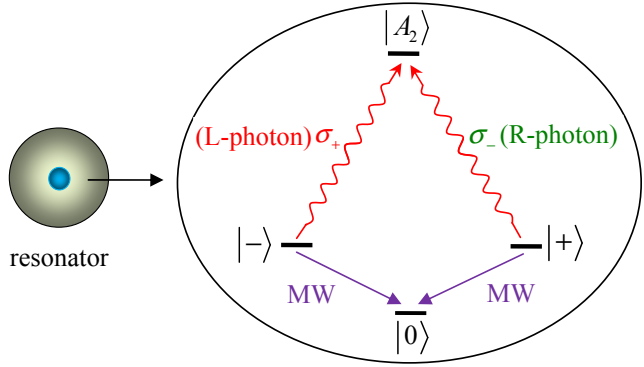}
\caption{A $\wedge$-type  atom-like structure of the
negatively-charged NV centre confined in an
optical resonator. The states  $|\pm\rangle$ act as the computational
basis states, and the state  $|0\rangle$ acts as an ancilla employed
for spin manipulation. The optical transitions from the ground
states $|\pm\rangle$ to the ancillary state $|A_2\rangle$ are
coupled by the $\sigma_{\mp}$ circularly polarized photons,
respectively.} \label{level}
\end{center}
\end{figure}

The NV centre consists of a substitutional nitrogen atom
replacing a carbon atom and an adjacent vacancy in the diamond
lattice. The ground states of the NV centre are an electronic
spin-triplet state with a 2.88-GHz zero-field splitting between the
magnetic sublevels with the angular momentum $m_s=0$ and $m_s=\pm1$
resulting from the spin-spin interactions \cite{split}. The
 states $|m_s=\pm1\rangle$ (denoted by $|\pm\rangle$) are coupled to
the state   $|m_s=0\rangle$ (denoted by $|0\rangle$) with microwave
pulses $\Omega_{\pm}$ forming a $\vee$-type three-level system.  At low temperature (T$\approx$4 K), the  states  $|\pm\rangle$ are coupled to one of the six excited states \cite{group}
$|A_2\rangle=(|E_-\rangle|+\rangle+|E_+\rangle|-\rangle)/\sqrt{2}$
with optical radiation forming a $\wedge$-type system [see Fig. \ref{level}].
The transitions $|\pm\rangle\rightarrow|A_2\rangle$ are driven by
the $\sigma_{\mp}$ polarized single photons at $\sim$ 637 nm,
respectively, and then  the state $|A_2\rangle$ spontaneously decays
into the states $|\pm\rangle$ with equal probabilities. Here
$|E_{\pm}\rangle$ indicate the states with the angular momentum
$\pm1$ along the NV centre axis, respectively.  The state $|A_2\rangle$, an
inherent spin-orbit entangled state and are protected by the spin-orbit
and spin-spin interactions, is robust against the small strain  and
magnetic fields, preserving the polarization properties of its
optical transition \cite{split1}.

A diagram of a basic spin-photon entangling unit,  a $\wedge$-type
three-level NV centre trapped in a single-sided cavity, is
shown in Fig. \ref{level}. An
incident single photon with frequency $\omega_p$ enters a
single-sided cavity with frequency $\omega_c$. The
cavity mode $\hat{a}$ with the right and left circular polarizations, $R$ and $L$, are
resonantly coupled to the transitions
$|\pm\rangle\rightarrow|A_2\rangle$  with frequency $\omega_0$, respectively. The cavity mode is driven by the input field
$\hat{a}_{in}$. By solving the Heisenberg equations of motion for
the annihilation operation $\hat{a}$ of cavity mode and the lowing
operation $\sigma_-$ of the NV centre \cite{QObook},
\begin{eqnarray}       \label{eq1}
&&\frac{d\hat{a}}{dt}  = -\left[i(\omega_c-\omega_p)+\frac{\kappa}{2}\right]\hat{a}(t)-\textrm{g}\sigma_{-}(t) - \sqrt{\kappa}\hat{a}_{in}, \nonumber\\
&&\frac{d\sigma_-}{dt} =
-\left[i(\omega_{0}-\omega_p)+\frac{\gamma}{2}\right]\sigma_{-}(t)-\textrm{g}\sigma_z(t)\hat{a}(t)+ \sqrt{\gamma}\sigma_z(t)\hat{b}_{in}(t),
\end{eqnarray}
one \cite{NV-NV3,Hu2008} can obtain the reflection coefficient for the
NV-centre-cavity unit in the weak excitation limit
$\langle\sigma_z\rangle=-1$,
\begin{eqnarray}       \label{eq2}
r(\omega_p)=\frac{\hat{a}_{out}}{\hat{a}_{in}}=\frac{[i(\omega_{c}-\omega_p)-\frac{\kappa}{2}][i(\omega_{0}-\omega_p)+\frac{\gamma}{2}]+\textrm{g}^2}
           {[i(\omega_{c}-\omega_p)+\frac{\kappa}{2}][i(\omega_{0}-\omega_p)+\frac{\gamma}{2}]+\textrm{g}^2}.
\end{eqnarray}
Here, the cavity output field $\hat{a}_{out}$ is connected with the
input field $\hat{a}_{in}$ by the input-output relation
$\hat{a}_{out} = \hat{a}_{in}+ \sqrt{\kappa}\hat{a}(t)$.
$b_{in}(t)$, as the vacuum input field, has the standard commutation
relation $[\hat{b}_{in}(t),\hat{b}_{in}^\dag(t')]=\delta(t-t')$.
$\gamma$ is the decay rate of the NV centre population.
$\kappa$ is the damping rate of the cavity intensity. $\sigma_z$ is
the inversion operator of the NV centre. $\textrm{g}$ is the coupling rate of the
NV-centre-cavity system.

From Eq. (\ref{eq2}), one can see that if the single photon feels a
hot cavity ($\textrm{g}\neq0$), after reflection, it will get a phase
shift $e^{i\varphi}$ with amplitude $|r(\omega_p)|$. If the photon
feels a cold cavity ($\textrm{g}=0$), after reflection, it will
acquire a phase shift $e^{i\varphi_0}$ with amplitude
$|r_0(\omega_p)|$. Considering the NV centre is prepared in the state
$|+\rangle$, the $R$- ($L$-) polarized photon feels a hot (cold) cavity, and the corresponding state
transformations are
\begin{eqnarray}       \label{eq3}
&&|R\rangle|+\rangle \;\rightarrow\; r(\omega_p)|R\rangle|+\rangle=e^{i\varphi}|r(\omega_p)||R\rangle|+\rangle,\nonumber\\
&&|L\rangle|+\rangle \;\rightarrow\;
r_0(\omega_p)|L\rangle|+\rangle=e^{i\varphi_0}|r_0(\omega_p)||L\rangle|+\rangle.
\end{eqnarray}
In case the NV centre electron spin is in the state $|-\rangle$, the corresponding transformations are
\begin{eqnarray}       \label{eq4}
&&|R\rangle|-\rangle \;\rightarrow\; r_0(\omega_p)|R\rangle|-\rangle=e^{i\varphi_0}|r_0(\omega_p)||R\rangle|-\rangle,\nonumber\\
&&|L\rangle|-\rangle \;\rightarrow\;r(\omega_p)|L\rangle|-\rangle=e^{i\varphi}|r(\omega_p)||L\rangle|-\rangle.
\end{eqnarray}
Here $r_0$ is described by  Eq. (\ref{eq2}) with $\textrm{g}=0$. The
conditional phase shifts are the functions of the frequency detuning
($\omega_p-\omega_c$) under the resonant condition
$\omega_c=\omega_0$. By adjusting
$\omega_p=\omega_c=\omega_0$, it is  possible to reach
\begin{eqnarray}       \label{eq5}
r=\frac{-\kappa\gamma+4\textrm{g}^2}{\kappa\gamma+4\textrm{g}^2},\qquad\qquad
r_0=-1.
\end{eqnarray}
Ref. \cite{NV-NV3} shows that when
$\textrm{g}\geq5\sqrt{\gamma\kappa}$, $r(\omega_p)\simeq1$. That is, the reflection of an uncoupled single
photon can result in a phase shift $\pi$  between the NV centre electron
spin and the photon relative to the coupled one. The sign change of
the reflected single photon can be specifically summarized as:
\begin{eqnarray}       \label{eq6}
&&|R\rangle|+\rangle \;\;\rightarrow\;\;|R\rangle|+\rangle, \;\;\;\;\;\;\;\;\;
|R\rangle|-\rangle\;\; \rightarrow\;\;-|R\rangle|-\rangle,\nonumber\\
&&|L\rangle|+\rangle\;\;\rightarrow\;\;-|L\rangle|+\rangle,\;\;\;\;\;\;\;
|L\rangle|-\rangle \;\;\rightarrow\;\;|L\rangle|-\rangle.
\end{eqnarray}

The emitted photon-matter entangling platform is at the heart of the photon-based or matter-based scalable quantum computing \cite{ourgatesNV,ourphotongate}, Bell-state analysis \cite{Bell-analisis},  entanglement detection \cite{detection}, and it has been received great progress in experiment. In 2008, Fushman \emph{et al.} \cite{emitted1} observed phase shift $\pi/4$ in QD-cavity system. In 2014, Reiserer \emph{et al.} \cite{Atom-CPF1} and Tiecke \emph{et al.} \cite{Atom-CPF2} observed phase shift $\pi$ in atom-cavity system. In 2016, Sun \emph{et al.} \cite{emitted2} observed phase shift $\pi$ in QD-cavity system. In the following, we exploit the spin-selective optical transition
property described by Eq. (\ref{eq6}) to study the implementation of the multi-photon
hyper-parallel quantum computing.

\begin{figure}[!h]
\begin{center}
\includegraphics[width=13.0 cm,angle=0]{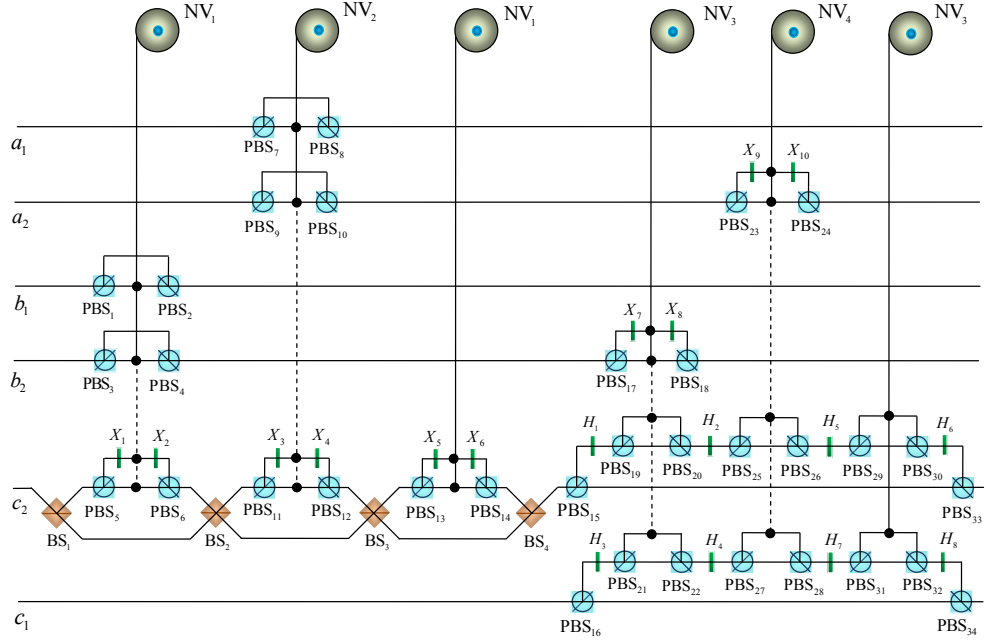}
\caption{Compact quantum circuit for implementing
hyper-parallel $\rm{C^2PF}$ gate on a three-photon system with  both
the spatial and polarization DOFs. CPBS$_i$ ($i=1,\;2,\;\cdots,34$)
represents a circular polarizing beam splitter  that transmits the
right circular polarizations ($R$) and reflects the left circular
polarizations ($L$). $H_i$ ($i=1,\;2,\;\cdots,8$) represents a
Hadamard operation performed on the passing photon with a half-wave
plate oriented at 22.5$^\circ$. $X_j$ ($j=1,\;2,\;\cdots,10$)
represents a bit-flip operation performed on the passing photon with a half-wave
plate oriented at 45$^\circ$. BS$_k$ ($k=1,\;2,\;3,\;4$) is a
balanced polarization-preserving beam splitter.}
\label{Toffoli}
\end{center}
\end{figure}

Figure \ref{Toffoli} depicts a compact quantum circuit for determinately
implementing a hyper-$\rm{C^2PF}$ gate acting on a three-photon
system in both the polarization and spatial DOFs. This gate
independently operates two $\rm{C^2PF}$ gates simultaneously on a
three-photon system. The $\rm{C^2PF}$ gate is equivalent to the
Toffoli gate up to two Hadamard transformations, i.e.,
$U_{\rm{C^2PF}}=(I_4\otimes H)U_{\rm{Toffoli}}(I_4\otimes H)$. Here
$I_4$ is a $4\times4$ identity matrix and $H$ represents the
Hadamard operation performed on the target qubit. Let us assume  that the single photons $a$, $b$,
and $c$ are initially prepared in the normalized states
\begin{eqnarray}       \label{eq7}
&&|\varphi\rangle_{a}=(\alpha_1|R_1\rangle+\alpha_2|L_1\rangle)\otimes(\varsigma_1|a_1\rangle+\varsigma_2|a_2\rangle),\nonumber\\
&&|\varphi\rangle_{b}=(\beta_1|R_2\rangle+\beta_2|L_2\rangle)\otimes(\zeta_1|b_1\rangle+\zeta_2|b_2\rangle),\nonumber\\
&&|\varphi\rangle_{c}=(\delta_1|R_3\rangle+\delta_2|L_3\rangle)\otimes(\xi_1|c_1\rangle+\xi_2|c_2\rangle).
\end{eqnarray}
The four NV centres, NV$_{1,2,3,4}$, with nearly identical
electron-spin energy levels are respectively prepared in the states
\begin{eqnarray}       \label{eq8}
|\varphi\rangle_{e_{1,3}}=\frac{|+_{1,3}\rangle+|-_{1,3}\rangle}{\sqrt{2}},\;\;\;\;
|\varphi\rangle_{e_{2,4}}=\frac{|+_{2,4}\rangle-|-_{2,4}\rangle}{\sqrt{2}}.
\end{eqnarray}
Here, $a_1$ ($b_1$ or $c_1$) and $a_2$ ($b_2$ or $c_2$) are the two
spatial modes of the photon $a$ ($b$ or $c$). $R_1$ ($L_1$), $R_2$
($L_2$), and $R_3$ ($L_3$) denote the photon $a$, $b$, or $c$ are in
the right (left) circular polarization states, respectively. The subscript $j$ of the $|+_j\rangle$ ($|-_j\rangle$) represents the $j$-th NV centre is in the state $|+\rangle$ ($|-\rangle$).

Now, let us go into the detail of our scheme in step by step for
implementing the hyper-$\rm{C^2PF}$ gate.

First, photon $b$  passes through the block [$\textrm{PBS}_1\rightarrow \textrm{NV}_1 \rightarrow\textrm{PBS}_2$ or $\textrm{PBS}_3\rightarrow \textrm{NV}_1 \rightarrow\textrm{PBS}_4$]. Subsequently, a Hadamard
transformation $H_{\rm{NV}_1}$ is performed on NV$_1$ by applying a
$\pi/2$ microwave pulse, which completes the transformations:
 \begin{eqnarray}       \label{eq9}
|+\rangle\rightarrow\frac{1}{\sqrt{2}}(|+\rangle+|-\rangle),\;\;\;\;|-\rangle\rightarrow\frac{1}{\sqrt{2}}(|+\rangle-|-\rangle).
\end{eqnarray}
Here, PBS$_i$ ($i=1,\dots,4$) is the circularly
polarizing beam splitter which lets the component  $R$ of the
incident photon be transmitted, while having the component $L$ be
reflected. The operations ($\textrm{PBS}_1\rightarrow \textrm{NV}_1 \rightarrow\textrm{PBS}_2 \rightarrow H_{\rm{NV}_1}$ and $\textrm{PBS}_3\rightarrow \textrm{NV}_1 \rightarrow\textrm{PBS}_4 \rightarrow H_{\rm{NV}_1}$)
transform the total state of the system composed of photons $a$,
$b$, $c$, and NV$_{1,2,3,4}$ from $|\varphi_0\rangle$
to $|\varphi_1\rangle$. Here
\begin{eqnarray}       \label{eq10}
|\varphi_0\rangle=|\varphi\rangle_{a}\otimes|\varphi\rangle_{b}\otimes|\varphi\rangle_{c}
\otimes|\varphi\rangle_{e_1}\otimes|\varphi\rangle_{e_2}\otimes|\varphi\rangle_{e_3}\otimes|\varphi\rangle_{e_4},
\end{eqnarray}
\begin{eqnarray}       \label{eq11}
|\varphi_1\rangle&=&|\varphi\rangle_{a}\otimes|\varphi\rangle_{c}\otimes|\varphi\rangle_{e_2}
\otimes|\varphi\rangle_{e_3}\otimes|\varphi\rangle_{e_4}\otimes(\zeta_1|b_1\rangle+\zeta_2|b_2\rangle)\nonumber\\
&&\otimes(\beta_1|R_2\rangle|-_1\rangle+\beta_2|L_2\rangle|+_1\rangle).
\end{eqnarray}

Second, photon $c$ emitting from the spatial mode $c_2$ passes through
the block [$\textrm{BS}_1\rightarrow\textrm{PBS}_{5}\rightarrow
X_1\rightarrow \textrm{NV}_1\rightarrow X_2\rightarrow
\rm{PBS}_{6}\rightarrow \rm{BS}_2$]. These operations make
$|\varphi_1\rangle$ be changed into
\begin{eqnarray}       \label{eq12}
|\varphi_2\rangle&=&|\varphi\rangle_{a}
\otimes|\varphi\rangle_{e_2}
\otimes|\varphi\rangle_{e_3}
\otimes|\varphi\rangle_{e_4}
\otimes(\zeta_1|b_1\rangle+\zeta_2|b_2\rangle)
\otimes(\delta_1|R_3\rangle+\delta_2|L_3\rangle)\nonumber\\&&
\otimes
\big[\beta_1|R_2\rangle(\xi_1|c_1\rangle-\xi_2|c_{3}\rangle)|-_1\rangle+\beta_2|L_2\rangle(\xi_1|c_1\rangle+\xi_2|c_{2}\rangle)|+_1\rangle\big].
\end{eqnarray}
Here, $X_{1,2}$ represent the bit-flip operations performed on the
passing photon with the half-wave plates oriented at 45$^\circ$, i.e.,
$|R\rangle\leftrightarrow|L\rangle$.  The balanced nonpolarizing beam splitter (BS), say BS$_1$
(BS$_2$), on the spatial states $|c_2\rangle$ and $|c_3\rangle$ is defined
as
\begin{eqnarray}       \label{eq13}
|c_2\rangle\rightarrow\frac{1}{\sqrt{2}}(|c_2\rangle+|c_3\rangle),\;\;\;\;
|c_3\rangle\rightarrow\frac{1}{\sqrt{2}}(|c_2\rangle-|c_3\rangle).
\end{eqnarray}

Third, photon $a$ is injected into the block [$\textrm{PBS}_{7}\rightarrow \textrm{NV}_2 \rightarrow\textrm{PBS}_{8}$ or $\textrm{PBS}_{9}\rightarrow \textrm{NV}_2 \rightarrow\textrm{PBS}_{10}$], followed with an
$H_{\rm{NV}_2}$ performed on NV$_2$. These operations
($\textrm{PBS}_{7}\rightarrow \textrm{NV}_2 \rightarrow\textrm{PBS}_{8} \rightarrow H_{\rm{NV}_2}$ and $\textrm{PBS}_{9}\rightarrow \textrm{NV}_2 \rightarrow\textrm{PBS}_{10} \rightarrow H_{\rm{NV}_2}$) transform $|\varphi_2\rangle$ into
\begin{eqnarray}       \label{eq14}
|\varphi_3\rangle&=& (\varsigma_1|a_1\rangle+\varsigma_2|a_2\rangle)
\otimes(\zeta_1|b_1\rangle+\zeta_2|b_2\rangle)
\otimes(\delta_1|R_3\rangle+\delta_2|L_3\rangle)
\nonumber\\&&\otimes(\alpha_1|R_1\rangle|+_2\rangle+\alpha_2|L_1\rangle|-_2\rangle)
\otimes
\big[\beta_1|R_2\rangle(\xi_1|c_1\rangle-\xi_2|c_{3}\rangle)|-_1\rangle\nonumber\\&&+\beta_2|L_2\rangle(\xi_1|c_1\rangle+\xi_2|c_{2}\rangle)|+_1\rangle\big]
\otimes|\varphi\rangle_{e_3} \otimes|\varphi\rangle_{e_4}.
\end{eqnarray}

Fourth,  photon $c$ emitting from the spatial mode $c_2$ or $c_3$ passes
through the block [$\textrm{PBS}_{11}\rightarrow X_3\rightarrow
\textrm{NV}_2\rightarrow X_4\rightarrow \textrm{PBS}_{12}$] and the
block [$\textrm{BS}_3\rightarrow\textrm{PBS}_{13}\rightarrow
X_5\rightarrow \textrm{NV}_1\rightarrow X_{6}\rightarrow
\textrm{PBS}_{14}\rightarrow\textrm{BS}_4$] successively. After these
two blocks, the state of the whole system becomes
\begin{eqnarray}       \label{eq15}
|\varphi_4\rangle&=&(\varsigma_1|a_1\rangle+\varsigma_2|a_2\rangle)
\otimes(\zeta_1|b_1\rangle+\zeta_2|b_2\rangle)
\otimes(\delta_1|R_3\rangle+\delta_2|L_3\rangle)
\nonumber\\&&\otimes\big\{\alpha_1|R_1\rangle|+_2\rangle(\beta_1|R_2\rangle|-_1\rangle+\beta_2|L_2\rangle|+_1\rangle)(\xi_1|c_1\rangle+\xi_2|c_{2}\rangle)
\nonumber\\
&&+\alpha_2|L_1\rangle|-_2\rangle\big[\beta_1|R_2\rangle|-_1\rangle(\xi_1|c_1\rangle+\xi_2|c_{2}\rangle)
\nonumber\\&&+\beta_2|L_2\rangle|+_1\rangle(\xi_1|c_1\rangle-\xi_2|c_{2}\rangle)\big]
\big\}\otimes|\varphi\rangle_{e_3} \otimes|\varphi\rangle_{e_4}.
\end{eqnarray}


Fifth, photon $b$ in the spatial mode $b_2$ passes through the block
[$\textrm{PBS}_{17}\rightarrow X_7\rightarrow\textrm{NV}_3\rightarrow
X_8\rightarrow\textrm{PBS}_{18}$], followed by an $H_{\rm{NV}_3}$
performed on NV$_3$. The above operations
($\textrm{PBS}_{17}\rightarrow X_7\rightarrow\textrm{NV}_3\rightarrow
X_8\rightarrow\textrm{PBS}_{18}\rightarrow H_{\rm{NV}_3}$) transform
$|\varphi_4\rangle$ into
\begin{eqnarray}       \label{eq16}
|\varphi_5\rangle&=&
\big\{\alpha_1|R_1\rangle|+_2\rangle(\beta_1|R_2\rangle|-_1\rangle+\beta_2|L_2\rangle|+_1\rangle)(\xi_1|c_1\rangle+\xi_2|c_{2}\rangle)
\nonumber\\
&&+\alpha_2|L_1\rangle|-_2\rangle\big[\beta_1|R_2\rangle|-_1\rangle(\xi_1|c_1\rangle+\xi_2|c_{2}\rangle)
\nonumber\\&&+\beta_2|L_2\rangle|+_1\rangle(\xi_1|c_1\rangle-\xi_2|c_{2}\rangle)\big]
\big\} \otimes(\varsigma_1|a_1\rangle+\varsigma_2|a_2\rangle)\nonumber\\&&
\otimes(\zeta_1|b_1\rangle|+_3\rangle+\zeta_2|b_2\rangle|-_3\rangle)
\otimes(\delta_1|R_3\rangle+\delta_2|L_3\rangle)
\otimes|\varphi\rangle_{e_4}.
\end{eqnarray}

Sixth, after photon $c$ passes through CPBS$_{15}$ (CPBS$_{16}$), the
$L$-polarized component passes through the block [$ \textrm{PBS}_{19}\rightarrow \textrm{NV}_3 \rightarrow \textrm{PBS}_{20}$] or the block [$\textrm{PBS}_{21}\rightarrow \textrm{NV}_3\rightarrow \textrm{PBS}_{22}$]. Before and after the
wave-packet interacts with such two blocks, a Hadamard operation $H_p$, which
completes the transformations
$|R\rangle\rightarrow(|R\rangle+|L\rangle)/\sqrt{2}$ and
$|L\rangle\rightarrow(|R\rangle-|L\rangle)/\sqrt{2}$, is applied on
it with the half  wave-plates $H_1$ and $H_2$ ($H_3$ and $H_4$)
oriented at 22.5$^\circ$, respectively. The operations
($\textrm{CPBS}_{15}  \rightarrow H_1 \rightarrow \textrm{PBS}_{19} \rightarrow \textrm{NV}_3 \rightarrow \textrm{PBS}_{20} \rightarrow H_2$ and $\textrm{CPBS}_{16}  \rightarrow H_3 \rightarrow \textrm{PBS}_{21} \rightarrow \textrm{NV}_3 \rightarrow \textrm{PBS}_{22} \rightarrow H_4$) transform $|\varphi_5\rangle$ into
\begin{eqnarray}       \label{eq17}
|\varphi_6\rangle&=&
\big\{\alpha_1|R_1\rangle|+_2\rangle(\beta_1|R_2\rangle|-_1\rangle+\beta_2|L_2\rangle|+_1\rangle)(\xi_1|c_1\rangle+\xi_2|c_{2}\rangle)
\nonumber\\
&&+\alpha_2|L_1\rangle|-_2\rangle\big[\beta_1|R_2\rangle|-_1\rangle(\xi_1|c_1\rangle+\xi_2|c_{2}\rangle)
+\beta_2|L_2\rangle|+_1\rangle\nonumber\\&&(\xi_1|c_1\rangle-\xi_2|c_{2}\rangle)\big]
\big\}
\otimes(\varsigma_1|a_1\rangle+\varsigma_2|a_2\rangle)
\big[\zeta_1|b_1\rangle(\delta_1|R_3\rangle+\delta_2|R_3\rangle)|+_3\rangle\nonumber\\&&+\zeta_2|b_2\rangle(\delta_1|R_3\rangle+\delta_2|L_3\rangle)|-_3\rangle
\big] \otimes|\varphi\rangle_{e_4}.
\end{eqnarray}

Seventh, photon $a$ in spatial mode $a_2$ passes through the block
[$\textrm{CPBS}_{23}\rightarrow X_9\rightarrow\textrm{NV}_4\rightarrow
X_{10}\rightarrow\textrm{CPBS}_{24}$]. Subsequently, an
$H_{\rm{NV}_4}$ is applied on NV$_4$. These operations transform
$|\varphi_6\rangle$ into
\begin{eqnarray}       \label{eq18}
|\varphi_7\rangle&=&
\big\{\alpha_1|R_1\rangle|+_2\rangle(\beta_1|R_2\rangle|-_1\rangle+\beta_2|L_2\rangle|+_1\rangle)(\xi_1|c_1\rangle+\xi_2|c_{2}\rangle)
\nonumber\\
&&+\alpha_2|L_1\rangle|-_2\rangle\big[\beta_1|R_2\rangle|-_1\rangle(\xi_1|c_1\rangle+\xi_2|c_{2}\rangle)
\nonumber\\&&+\beta_2|L_2\rangle|+_1\rangle(\xi_1|c_1\rangle-\xi_2|c_{2}\rangle)\big]
\big\}
\otimes(\varsigma_1|a_1\rangle|-_4\rangle+\varsigma_2|a_2\rangle|+_4\rangle)\nonumber\\
&&\otimes\big[\zeta_1|b_1\rangle(\delta_1|R_3\rangle+\delta_2|R_3\rangle)|+_3\rangle+\zeta_2|b_2\rangle(\delta_1|R_3\rangle+\delta_2|L_3\rangle)|-_3\rangle
\big].
\end{eqnarray}

Eighth, photon $c$ passes through the blocks [$\textrm{PBS}_{25} \rightarrow \textrm{NV}_{4} \rightarrow \textrm{PBS}_{26} \rightarrow H_5 \rightarrow \textrm{PBS}_{29} \rightarrow \textrm{NV}_{3} \rightarrow \textrm{PBS}_{30} \rightarrow H_6$] or the blocks [$\textrm{PBS}_{27} \rightarrow \textrm{NV}_{4} \rightarrow \textrm{PBS}_{28} \rightarrow H_7 \rightarrow \textrm{PBS}_{31} \rightarrow \textrm{NV}_{3} \rightarrow \textrm{PBS}_{32} \rightarrow H_8$]
successively, and the wave-packets of the photon $c$ are mixed at
PBS$_{33}$ (PBS$_{34}$).  The above operations change the state of the whole
system to be
\begin{eqnarray}       \label{eq19}
|\varphi_8\rangle&=&
\big\{\alpha_1|R_1\rangle|+_2\rangle\big[(\beta_1|R_2\rangle|-_1\rangle+\beta_2|L_2\rangle|+_1\rangle)(\xi_1|c_1\rangle+\xi_2|c_{2}\rangle)\big]
\nonumber\\
&&+\alpha_2|L_1\rangle|-_2\rangle\big[\beta_1|R_2\rangle(\xi_1|c_1\rangle+\xi_2|c_{2}\rangle)|-_1\rangle\nonumber\\&&+\beta_2|L_2\rangle(\xi_1|c_1\rangle-\xi_2|c_{2}\rangle)|+_1\rangle\big]
\big\}\nonumber\\&&
\otimes\{\varsigma_1|a_1\rangle|-_4\rangle[(\zeta_1|b_1\rangle|+_3\rangle+\zeta_2|b_2\rangle|-_3\rangle)(\delta_1|R_3\rangle+\delta_2|L_3\rangle)
\big]\nonumber\\&&
+\varsigma_2|a_2\rangle|+_4\rangle[\zeta_1|b_1\rangle(\delta_1|R_3\rangle+\delta_2|L_3\rangle)|+_3\rangle\nonumber\\&&
+\zeta_2|b_2\rangle(\delta_1|R_3\rangle-\delta_2|L_3\rangle)|-_3\rangle
\big]\}.
\end{eqnarray}

Finally, we measure the spins of NV$_{1,2,3,4}$ in the basis
$\{|\pm'\rangle=(|+\rangle\pm|-\rangle)/\sqrt{2}\}$ to disentangle
the NV centres.  Subsequently, the classical feed-forward single-qubit
operations, as shown in Tab. \ref{tableToffoli}, are performed on
the three outing photons to raise the success probability of our
hyper-${\rm{C^2PF}}$ gate to 100\% in principle. These operations
result in a three-photon output state
\begin{eqnarray}       \label{eq20}
|\varphi_9\rangle&=&
\big\{(\alpha_1\beta_1|R_1\rangle|R_2\rangle+\alpha_1\beta_2|R_1\rangle|L_2\rangle+\alpha_2\beta_1|L_1\rangle|R_2\rangle)(\xi_1|c_1\rangle+\xi_2|c_{2}\rangle)
\nonumber\\
&&+\alpha_2\beta_2|L_1\rangle|L_2\rangle(\xi_1|c_1\rangle-\xi_2|c_{2}\rangle)
\big\}\nonumber\\
&&
\otimes\big\{(\varsigma_1\zeta_1|a_1\rangle|b_1\rangle+\varsigma_1\zeta_2|a_1\rangle|b_2\rangle+\varsigma_2\zeta_1|a_2\rangle|b_1\rangle)
(\delta_1|R_3\rangle+\delta_2|L_3\rangle)\nonumber\\&&+\varsigma_2\zeta_2|a_2\rangle|b_2\rangle(\delta_1|R_3\rangle-\delta_2|L_3\rangle)
\big\}.
\end{eqnarray}
Therefore, the total operations for a deterministic
hyper-${\rm{C^2PF}}$ gate are completed. That is, the quantum circuit
shown in Fig. \ref{Toffoli} completes a hyper-parallel optical ${\rm{C^2PF}}$
gate which independently changes the phase of the input states by
$\pi$, that is, a sign change, if all qubits are in the states
$|L_1L_2c_2\rangle$ or $|a_2b_2L_3\rangle$, and has no effect
otherwise.

\begin{table}[htb]
\centering \caption{The classical feed-forward operations on the photonic
qubits to complete a full and deterministic hyper-$\rm{C^2PF}$ gate
conditioned on the outcomes of the NV centre spins.
 $\sigma_z=|R\rangle\langle R|-|L\rangle\langle L|$. Phase shifter $\pi$ performed on the spatial mode $a_1$ ($b_2$) completes the transformation $|a_1\rangle\rightarrow -|a_1\rangle$ ($|b_2\rangle\rightarrow -|b_2\rangle$).}
\begin{tabular}{cc}
\hline  \hline

 NV centres  &  classical feed-forward operation \\ \hline
 $|-_4'\rangle$ ($|+_4'\rangle$) &  phase shift $\pi$ is performed on the spatial mode $a_1$ (no)\\ \hline

 $|-_3'\rangle$ ($|+_3'\rangle$) &  phase shift $\pi$ is performed on the spatial mode $b_2$ (no) \\ \hline

 $|-_2'\rangle$ ($|+_2'\rangle$) &  $\sigma_z$ is performed on photon 1 (no)\\ \hline

 $|-_1'\rangle$ ($|+_1'\rangle$) &  $-\sigma_z$ is performed on photon 2 (no)\\ \hline

                             \hline\hline
\end{tabular}\label{tableToffoli}
\end{table}

\section{The average fidelity of the hyper-$\rm{C^2PF}$ gate}

\begin{figure}[!h]
\begin{center}
\includegraphics[width=6.2 cm,angle=0]{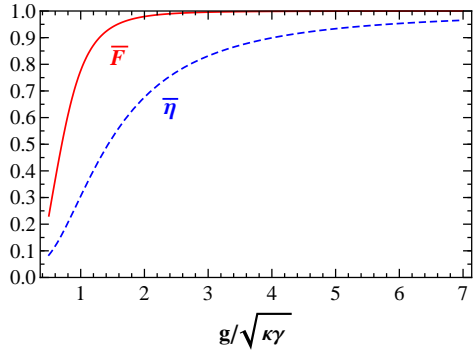}
\caption{ The average fidelities ($\overline{F}$, red
solid curve) and efficiencies ($\overline{\eta}$, blue dashed curve)
of the hyper-$\rm{C^2PF}$ gate as a function of
$\textrm{g}/\sqrt{\kappa\gamma}$.
$\textrm{g}/\sqrt{\kappa\gamma}\geq 0.5$.
} \label{Fidelity}
\end{center}
\end{figure}

The coefficients of the reflection photons, described by Eq.
(\ref{eq2}), play an important role in constructing our
hyper-$\rm{C^2PF}$ gate. The imperfection in phase and amplitude of
the reflection photons reduces the performance of our gate.
Therefore, it is necessary to consider the feasibility of our gate,
which can be evaluated by the fidelity of the final normalized
states in the realistic case $|\varphi_{\rm{out}}'\rangle$
relative to that in the ideal case $|\varphi_{\rm{out}}\rangle$
averaged  over all the input (output) states, that is,
\begin{eqnarray}       \label{eq22}
\overline{F}&=&\frac{1}{(2\pi)^6}\int_{0}^{2\pi}d\alpha\int_{0}^{2\pi}d\beta\int_{0}^{2\pi}d\gamma\int_{0}^{2\pi}
d\varsigma\int_{0}^{2\pi}d\zeta \int_{0}^{2\pi}d\xi\;   |\langle\psi_{\textrm{out}}|\psi_{\textrm{out}}'\rangle|^2.
\end{eqnarray}
Here $|\varphi_{\rm{out}}\rangle$ is described by Eq.
(\ref{eq19}). Using the same argument as for $|\varphi\rangle$,
$|\varphi_{\rm{out}}'\rangle$ can be obtained by substituting  Eqs.
(\ref{eq3})-(\ref{eq4}) for Eq. (\ref{eq6}).  Here, $\cos
\alpha=\alpha_1$,  $\sin \alpha=\alpha_2$, $\cos \beta=\beta_1$,
$\sin \beta=\beta_2$, $\cos \gamma=\gamma_1$, $\sin
\gamma=\gamma_2$, $\cos \varsigma=\varsigma_1$, $\sin
\varsigma=\varsigma_2$, $\cos \zeta=\zeta_1$, $\sin \zeta=\zeta_2$,
$\cos \xi=\xi_1$, and $\sin \xi=\xi_2$. By calculation, we find that
the average fidelity of our hyper-$\rm{C^2PF}$ gate, $\overline{F}$,
as a function of $\textrm{g}/\sqrt{\kappa\gamma}$ can be depicted by
the red solid curve in Fig. \ref{Fidelity}.

\section{The average efficiency of the hyper-$\rm{C^2PF}$ gate}
When the photon interacts with the matter qubit, the incident photon
is inevitably  lost. In order to show the feasibility of our gate,
it is also necessary to consider the efficiency of our gate, $\eta$,
which is defined as the yield of the incident photon, that is, the
ratio of the gate's output photon number $n_{\rm{output}}$ to its
input photon number $n_{\rm{input}}$. Taking all the possible input
(output) states, that is,  average over
$\alpha,\;\beta,\;\gamma,\;\varsigma,\;\zeta,\;\xi\;\in\;[0, 2\pi]$,
the average efficiency of our gate is given by
\begin{eqnarray}       \label{eq22}
\overline{\eta}&=&\frac{1}{(2\pi)^6}\int_{0}^{2\pi}d\alpha\int_{0}^{2\pi}d\beta\int_{0}^{2\pi}d\gamma\int_{0}^{2\pi}
d\varsigma\int_{0}^{2\pi}d\zeta \int_{0}^{2\pi}d\xi\; \frac{n_{\rm{output}}}{n_{\rm{input}}}\nonumber \\
&=& [121 +(128|r| + 164|r|^2 + 40|r|^3 + 14 |r|^4+ |r|^5 (2 + |r|)^2 (4 + |r|))\nonumber\\&& \times(91 +
   58|r| +42|r|^2 +18|r|^3 + 12|r|^4 + 14|r|^5 +14|r|^6  +|r|^7 (6 +|r|)]/131072.\nonumber\\
\end{eqnarray}
The dependence of the average efficiency  of our hyper-$\rm{C^2PF}$
gate, $\overline{\eta}$, on the coupling strength
$\textrm{g}/\sqrt{\kappa\gamma}$ is shown in Fig. \ref{Fidelity} (see the blue
dashed curve).

\section{The challenges of our scheme.}
In the work, the gate mechanism is deterministic in principle. However, the experimental imperfections degrade the gate fidelity and efficiency. The main sources of the errors
include the finite signal-to-noise ratio in the zero-phonon line (ZPL) channel (reduces
the fidelity of the platform by 11\%) \cite{Togan}, the excited-state
mixing caused by the strain and the
depolarization of the single photons (reduces the fidelity of the
platform by 12\%) \cite{Togan}, timing jitter (reduces the fidelity
of the platform by 4\%) \cite{Togan}, spin-flips during the optical
excitation result in the spin-spin interaction (probability
$0.46\pm0.01\%$ when T=4 K) \cite{three-meter}. There are
other errors due to the technical imperfections, such as imperfections in the NV centre electron spin
population into state $m_s=0$ (fidelity $99.7\pm0.1\%$) and
 states $m_s=\pm1$ (fidelity $99.2\pm0.1\%$) \cite{readout3}, the
imperfect rotations of the NV centre electron spin qubit by using
microwave, the detector dark counts and background counts during the measurement of the NV centre, the
spatial mismatch between the cavity and incident photon, the
balancing of PBSs (extinction ratio about 100: 1 in reflection,
1000: 1 in transmission \cite{Atom-CPF1}) and BSs, and the intracavity loss and linear optical elements loss.  The weak narrow-band ZPL emission at 637 nm is one the drawbacks of the applications of the NV centers, and about 70-80\% of the NV centre's fluorescence emission is emitted into the narrow-band ZPL even at room temperature due to the low photon-electron coupling \cite{ZPL}. The imperfections due to the technical imperfection will be largely improved with the further technical advances.

\section{Discussion and Summary}

Optical quantum information processing has been received great
attention and generally are focused on the traditionary ones, in
which the information is encoded in the polarization DOF of photons
only. At variance with the tradition ones with one DOF, quantum information
processing approach with multiple DOFs\cite{algorithm3,hyperentanglement-analysis} is much less
affected by its environment, reduces the quantum sources, or can simplify one-way quantum computing and quantum algorithm \cite{Simple-algorithm,one-way1,one-way3}. Nowadays, hyper-parallel quantum gates have been recognized as an
elementary element of quantum information processing. Some schemes for optical hyper-CNOT gates assisted by QDs or NV
centres have been proposed by Ren \emph{et al}.\cite{renLPL,renSR,renhypercnotpra}.  Wang \emph{et al}.\cite{wangtiejun1,wangtiejun2} designed some quantum circuits
for hyper-parallel universal quantum gates acting on hybrid
photon-matter systems.

In this work, we have designed a compact quantum circuit for
implementing an optical hyper-$\rm{C^2PF}$ gate  on a  three-photon
system in  both the polarization and spatial DOFs through NV-centre-cavity
interactions. Great efforts have been made to interact the NV centre with the photons. The single photon
coupling to an NV centre electron spin has been reported in recent
years\cite{Togan,Kosaka,England}. An NV centre coupled to a
fiber-based microcavity or microring resonator has been
demonstrated, respectively\cite{fiber-Albrecht,toroidal-Hausmann}. A
single NV centre coupled to a degenerate cavity mode, necessary for
our scheme, can be achieved by employing the H1 photonic crystal
cavity\cite{unpolarized-photon1}, fiber-based
microcavity\cite{fiber-Albrecht}, ring
microresonators\cite{toroidal-Hausmann}, or
micropillar\cite{unpolarized-pillar3}. The identically optical
transition energies of the four separated NV centres, required for
our scheme, can be achieved by applying controlled external electric
fields\cite{electricfield}. Our scheme is nearly free from spectral
diffusion and charge fluctuation\cite{free-diffusion} because of the
narrow linewidth (40 MHz) of the  state $|A_2\rangle$ \cite{Kosaka}.
The spectral diffusion, a hurdle for applications of the NV
centres, is induced by a fluctuating electrostatic environment
(usually caused by ionized impurities and charge traps) around the
 NV centre\cite{reason-diffusion}. A number of techniques
have been actively explored to reduce and eliminate the spectral
diffusion\cite{reduce-diffusion1,reduce-diffusion2,reduce-diffusion3}.

In summary, we have presented a compact quantum circuit for
implementing a three-photon hyper-parallel $\rm{C^2PF}$ gate with
both the polarization and spatial DOFs, assisted by NV-centre-cavity
interactions. Different from the traditional approach with encoding
the qubit in polarization DOF of photons only, we encode the qubits
in both the polarization and spatial DOFs of three photons, and
our gate reduces the quantum resources consumed in quantum information
processing by a half and the effect of the decoherence caused by the
noise channels. In contrast to linear-optics- or parity-check-based
procedures, our scheme does not require auxiliary single photons or
maximally entangled pairs of photons. The cost is six CPF gates for
our hyper-$\rm{C^2PF}$ gate. Given the current technology, our
scheme may be experimentally feasible with a high fidelity and
efficiency.

\section*{Acknowledgements}

This work is supported by the National Natural Science Foundation of
China (NSFC) (11547138, 11474026, 11175094 and 91221205), the Fundamental Research Funds
for the Central Universities (06500024 and 2015KJJCA01), the
National Basic Research Program of China (2011CB9216002). GLL is a member of the Center of Atomic and Molecular
Nanosciences, Tsinghua University.

\end{document}